\documentclass[%
reprint,
superscriptaddress,
amsmath,amssymb,
prl,
floatfix,
]{revtex4-1}

\usepackage{graphicx}
\usepackage{dcolumn}
\usepackage{bm}
\usepackage{hyperref}
\usepackage{gensymb}
\usepackage{blindtext}

\usepackage{caption}
\usepackage{subcaption}

\begin{document}
	\title[Sign inversion in the terahertz photoconductivity of single-walled carbon nanotube films]{Sign inversion in the terahertz photoconductivity of single-walled carbon nanotube films}
	\author{Peter Karlsen}
	\email{peterkarlsen88@gmail.com}
	\affiliation{%
		School of Physics, University of Exeter, Stocker Road, EX4 4QL, United Kingdom
	}%
	\author{Mikhail V. Shuba}
	\email{mikhail.shuba@gmail.com}
	\author{Polina P. Kuzhir}
	\affiliation{
		Institute for Nuclear Problems, Belarus State University, Bobruiskaya 11, 220050 Minsk, Belarus
	}%
	
	\affiliation{
		Tomsk State University, Lenin Avenue 36, 634050, Tomsk, Russia
	}%
	
	\author{Albert G. Nasibulin} 
	\affiliation{
		Skolkovo Institute of Science and Technology, Skolkovo Innovation Center, Building 3, Moscow 143026, Russia
	}%
	\affiliation{
		Department of Applied Physics, Aalto University, School of Science, P.O. Box 15100, FI-00076 Espoo, Finland
	}%
	
	\author{Patrizia Lamberti} 
	\affiliation{
		Department of Information and Electrical Engineering and Applied Mathematics, University of Salerno, Fisciano (SA), Italy
	}%
	\author{Euan Hendry}%
	\email{E.Hendry@exeter.ac.uk}
	\affiliation{%
		School of Physics, University of Exeter, Stocker Road, EX4 4QL, United Kingdom
	}%
	\date{\today}
	\begin{abstract}
		In recent years, there have been conflicting reports regarding the ultrafast photoconductive response of films of single walled carbon nanotubes (CNTs), which apparently exhibit photoconductivities that can differ even in sign. Here, we observe explicitly that the THz photoconductivity of CNT films is a highly variable quantity which correlates with the length of the CNTs, while the chirality distribution has little influence. Moreover, by comparing the photo-induced change in THz conductivity with heat-induced changes, we show that both occur primarily due to heat-generated modification of the Drude electron relaxation rate, resulting in a broadening of the plasmonic resonance present in finite-length metallic and doped semiconducting CNTs. This clarifies the nature of the photo-response of CNT films and demonstrates the need to carefully consider the geometry of the CNTs, specifically the length, when considering them for application in optoelectronic devices. 
	\end{abstract}
	
	\pacs{Valid PACS appear here}
	\maketitle
	
	The optical and electronic properties of single-walled carbon nanotubes (CNTs) have been intensely investigated for several decades due to their fascinating physical properties and potential for advanced applications \cite{Reich2008,Portnoi2008,He2014,Titova2015,Zubair2016}.
	Understanding the ultrafast dynamics of photoexcited charge-carriers in CNTs is critical due to their potential applications in photonics and optoelectronics \cite{Nanot2012,Avouris2007,Avouris2008,Leonard2009,Zhang2016,Docherty2014a}. For this reason, many groups have utilized time-resolved measurements to study the ultrafast response of CNTs due to optical photoexcitation, documenting, for example, the presence of excitons in photoexcited CNTs \cite{Soavi2016a,Wang2005a,Srivastava2008,Chmeliov2016a,Luo2015a,Haugen2014,Wang2005a}. 
	
	While visible pulses can detect the presence of excitons, THz pulses are ideal for probing low energy excitations such as free-carriers and plasmons, since each of these species have distinct features in the THz photoconductivity \cite{Ulbricht2011}. Thus, a proper understanding of the THz response of CNTs is key to understanding the ultrafast charge-carrier mechanisms in CNTs. Many groups have utilized optical pump - THz probe time-domain spectroscopy to investigate the ultrafast charge-carrier dynamics in CNTs \cite{Xu2009,Kampfrath2008,Beard2008,Perfetti2006,Luo2015a,Jensen2013a}; however, there are conflicting reports of the sign and frequency dependence of the observed photoconductivity. This discrepancy has led to wildly different interpretations and conclusions about the photoinduced THz response. For example, Xu et al. \cite{Xu2009} deduced that excitons are the dominant photogenerated species detected in these experiments, while Luo et al. \cite{Luo2015a} concluded the ultrafast THz response originates from transitions between exciton states. Beard et al. \cite{Beard2008} and, more recently Jensen et al. \cite{Jensen2013a}, have meanwhile concluded that free carriers are the dominant photoexcited species, an interpretation broadly shared by Kampfrath et al. \cite{Kampfrath2008,Perfetti2006}, with small-gap interband transitions also contributing to the THz response. While most of these measurements have been carried out on samples of mixed chirality (i.e. mixed semiconducting and metallic CNTs) it is important to note that Beard et al. \cite{Beard2008} found THz photoconductivities of samples containing 94\% semiconducting and 93\% metallic CNTs to be similar. Moreover, discrepancies persist even for nominally similar samples, with Luo et al. \cite{Luo2015a} and Xu et al. \cite{Xu2009} reporting a photoconductivity of different sign for samples of predominately small-diameter semiconducting CNTs.
	
	Since all of these groups have measured CNTs under similar excitation and preparation conditions, these discrepancies must originate from a difference in the measured samples themselves. The key to understanding these discrepancies lies in the observation of a broad peak in the THz conductivity of CNTs, observed for the first time, to our knowledge, in \cite{Bommeli1997}. While there has been some discussion regarding the nature of this resonance, with some groups proposing an interband transition of small-gap CNTs \cite{Ugawa1999,Borondics2006,Nishimura2007,Kampfrath2008}, more recent papers \cite{Shuba2012,Zhang2013,Morimoto2014a,Falk2017a} show clear evidence that it results from a localized plasmon in finite-length CNTs, which we denote the \textit{finite-length effect}, first proposed in \cite{Akima2006,Slepyan2006}. Theoretical modelling \cite{Slepyan2010a} and experimental observations \cite{Shuba2012,Zhang2013,Morimoto2014a,Falk2017a} substantiates the dominant role of the finite-length effect in the equilibrium THz response. Understanding the true origin of this THz resonance is also key to understanding the ultrafast charge-carrier dynamics of CNTs. However, due to the inherent difficulty in fabricating isolated CNT samples, most measurements have been carried out on mixtures of CNTs with various distributions in length, thickness, chirality and bundle-size, all fabricated using a variety of techniques \cite{Kim2003,Kampfrath2008,Xu2009,Bauhofer2009,Slepyan2010a,Shuba2012,Beard2008}.  
	
	In this paper we use optical pump - THz probe time-domain spectroscopy  to systematically investigate the influence of tube length and chirality on the THz photoconductivity of thin-films comprising single-walled CNTs. We observe explicitly that the THz photoconductivity of CNT films is a highly variable quantity which correlates with the length of the CNTs, while the chirality distribution (i.e. the relative concentration of metallic vs semiconducting tubes) has very little influence. Moreover, by comparing the photo-induced change in THz conductivity ($\Delta\sigma_{ph}$) to the change on heating from 10 K to 300 K ($\Delta\sigma_{heat}$), we show that both occur primarily due to the temperature-induced modification of Drude electron relaxation rate, which results in a broadening of the plasmonic resonance present in finite-length metallic and doped semiconducting CNTs.
	
	To study the influence of tube length and chirality, we prepared five types of films comprising CNTs in bundled form, where the average lengths of the CNT bundles and the chirality distributions of the films varies significantly, see table \ref{tab:samples}. The details of the sample preparation can be found in the supplementary material, section S1. 
	\begin{table}[t]
		\caption{Summary of our CNT films; thickness (D), length (L) (average length in parenthesis), diameter (d), and content of semiconducting (sem.) and metallic (met.) CNTs. Note that $l$-, $m$-, and $s$-CNT is short notation for \textit{long}-, \textit{medium}-, and \textit{short}-CNT, respectively.}
		\begin{ruledtabular}
			\begin{tabular}{r|r|r|r|r|r}
				\textbf{Sample} & \textbf{D (nm)}& \textbf{L ($\mu$m)}& \textbf{d (nm)} & \textbf{sem.}& \textbf{met.} \\ \hline
				sem-CNT & 500 & 0.1--1 & 0.8--1.2& 99\% & 1\%\\ \hline
				met-CNT & 500 & 0.1--1 & 0.8--1.2& 5\% & 95\%\\ \hline
				l-CNT & 55 & 2--100 (10) & 1.3--2& 66\% & 33\% \\ \hline
				m-CNT & 500 & 0.3--2 (1) & 0.8--1.2 & 66\% & 33\% \\ \hline
				s-CNT & 800 & $<0.3$ & 0.8--1.2 & 66\% & 33\%				
			\end{tabular}
		\end{ruledtabular}
		\label{tab:samples}
	\end{table}
	
	In order to observe the influence of the broad THz peak on the photoconductivity of CNTs, it is important to probe at or below the resonance frequency, which typically lies in the range 1-10 THz \cite{Shuba2012}. We carried out both transmission and photoconductivity measurements over the range ~0.2--1.5 THz, where THz pulses were incident normal to our samples. Transmission spectra were obtained using a simple time-domain spectrometer, where THz pulses were generated and detected by commercially available Photoconductive Antennas (PCAs) \cite{Fattinger1989a} from \href{http://www.batop.com/index.html}{Batop} using a 40 MHz, 1064 nm, femtosecond fibre-laser from \href{http://ekspla.com/}{Ekspla}. To investigate the photoexcited THz response of our samples, we employed a 100 fs, 1050 Hz repetition rate, 800 nm Ti:Sapphire amplified laser, where THz pulses were generated and detected by optical rectification \cite{Zhang1992} and electro-optic sampling \cite{Wu1997}, respectively, in 1mm thick ZnTe crystals. To photoexcite the sample, we again use 800 nm pulses, with fluences in the range of 0.7--15 $\mu$J/cm$^2$. By analysing the frequency dependent transmission amplitude and phase of a sample (see supplementary material section S2), we can determine its complex equilibrium effective conductivity, $\sigma(\nu)$, as in references \cite{Beard2008,Xu2009,Zhang2013}, where $\nu$ is the frequency. Similarly, by recording the difference in transmission, $\Delta{}E=E_{exc}-E$, between a photoexcited ($E_{exc}$) and unexcited sample ($E$), a complex photoconductivity $\Delta\sigma_{ph}(\nu,\Delta\tau)$ can be obtained as a function of pump-probe delay-time $\Delta\tau$ (again, see supplementary material section S2). To investigate the temperature dependence of the THz conductivity in the range 10--300 K, we employed a closed cycle helium cryostat (\href{http://www.arscryo.com/}{ARS})\cite{Ulbricht2011} with quartz windows. Note that the relatively narrow bandwidth of our measurements is determined by the transmission through this cryostat system. 
	
	\begin{figure}
		\centering
		\includegraphics[width=1\linewidth]{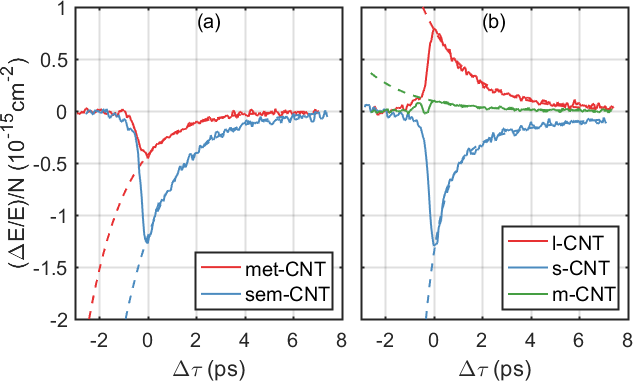}
		\caption{The photo-induced relative change in the THz transmission $\Delta{}E/E$ due to 800 nm photoexcitation at 300 K of \textbf{(a)} $met-$ and $sem-$CNT, and \textbf{(b)} $l-$, $m-$ and $s-$CNT, vs pump-probe delay time $\Delta\tau$ and normalized by the absorbed photon density $N$. The incident fluence is 15 $\mu$J/cm$^2$ for all films except $l-$CNT, where the fluence is 0.7 $\mu$J/cm$^2$. The full lines are the experimentally obtained data, and the dashed lines are exponential fits. The decay time $\tau$ is found to be 1.8 ps, 1.6 ps, 1.9 ps, and 1.9 ps for $sem-$, $met-$, $l-$, and $m-$, respectively. For $s-$CNT, an initial fast decay of 0.7 ps is observed, followed by a slow decay of 4.4 ps.}
		\label{fig:pumpvsl}
	\end{figure}
	\begin{figure}[tb]
		\centering
		\includegraphics[width=1\linewidth]{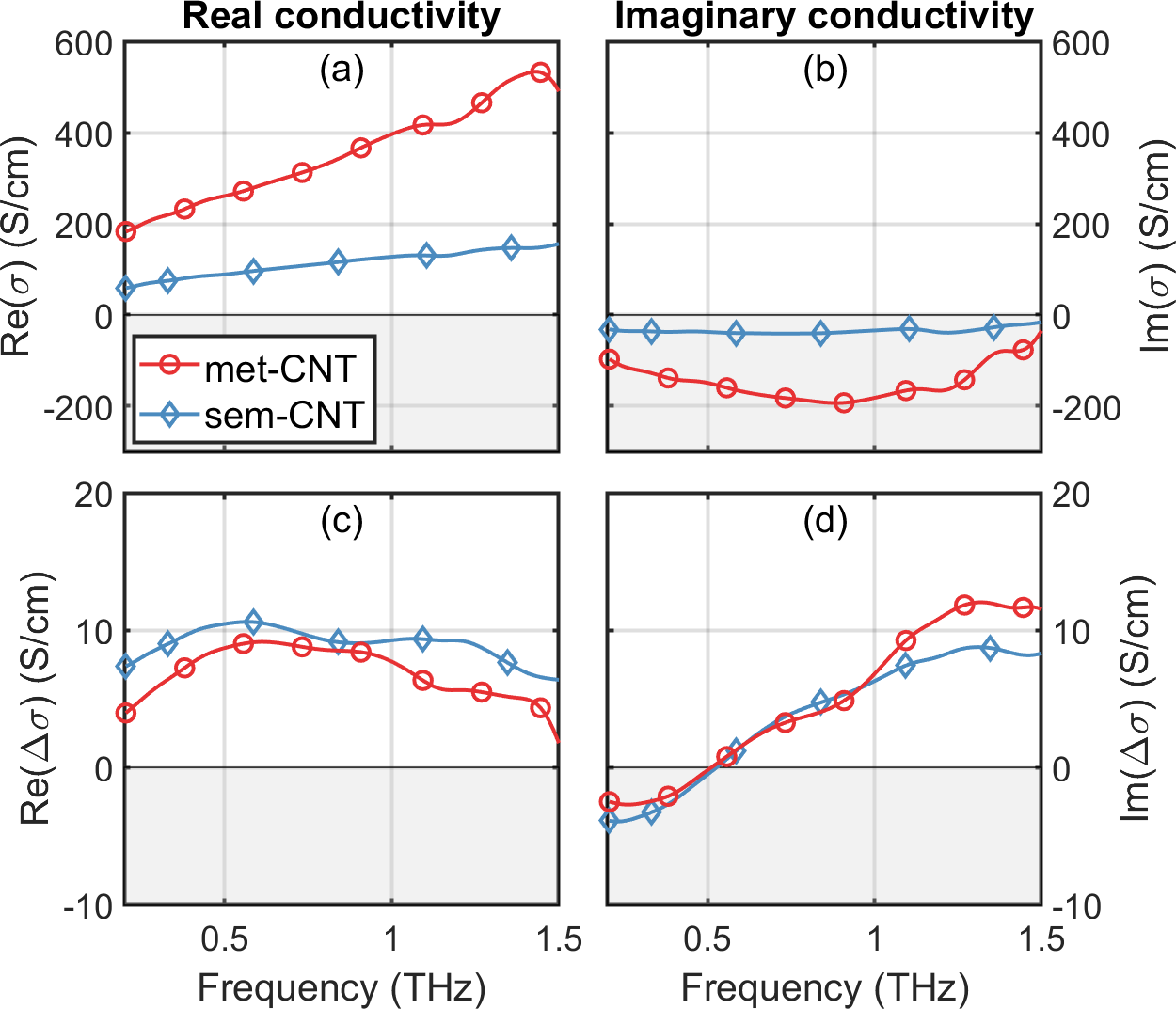}
		\caption{Effective conductivity \textbf{(a)} Re$(\sigma)$ and \textbf{(b)} Im$(\sigma)$ of $met-$ and $sem-$CNT at 300 K, and change in effective conductivity \textbf{(c)} Re$(\Delta\sigma)$ and \textbf{(d)} Im$(\Delta\sigma)$ due to 800 nm photoexcitation at pump-probe delay time $\Delta\tau=1$ ps. The incident fluence is 15 $\mu$J/cm$^2$. The negative region of the second axis in (a)-(d) have been shaded to highlight the difference in sign of $\sigma$ and $\Delta\sigma$.}
		\label{fig:condexcvsChirality}
	\end{figure}
	
	In figure \ref{fig:pumpvsl}, we plot the photoinduced change in transmission ($\Delta{}E/E$) as a function of pump-probe delay time $\Delta\tau$ and normalized to the absorbed photon density $N$, for the $met-$ and $sem-$CNT (a), and the $l-$, $m-$ and $s-$CNT films (b) \footnote{Note that in order to facilitate comparison of photoconductivities of similar order, the incident fluence $F$ used for the $l-$CNT is two orders of magnitude smaller than for the other samples, due to the relatively large photo-response of $l-$CNT (however, the dynamics and conductivity spectra are observed to be relatively fluence independent for our samples - see supplementary section S3)}. We note that the 800 nm photoexcitation occurs primarily off-resonance in terms of the optical transitions in the CNTs, meaning the only on-resonance photoexcitation occurs for a small subset of the semiconducting CNTs in the $s$- and $m$-CNTs, see supplementary material section S3.
	We observe that the decay dynamics of all the films are quite similar, with decay-times in the range of 1.6-1.9 ps, comparable to previous reports \cite{Lauret2003b,Wang2004,Jensen2013a,Xu2009,Perfetti2006}, which has previously been attributed to Auger recombination of the photoexcited electron-hole pairs \cite{Xu2009,Wang2004}.
	Since we observe little fluence dependence in the decay times of the various CNT films (see supplementary material section S3), we rule out Auger recombination as a significant relaxation mechanism for our films. Instead we associate the THz photoresponse and decay times with cooling of the CNT electronic system and lattice, which will become evident later on. While the decay dynamics of the different films are similar, the magnitude of $\Delta{}E/E$ varies significantly between the samples, and even changes sign. It is the origin of this large variation in photo response that forms the basis of this paper \footnote{The initial oscillatory behaviour of the $m$-CNT film is difficult to interpret, since it occurs on a sub-picosecond timescale, meaning it could very likely be an artefact from our measurement technique \cite{Nienhuys2005e}.}. 
	
	In figures \ref{fig:condexcvsChirality}a and \ref{fig:condexcvsChirality}b we plot the real and imaginary parts of $\sigma(\nu)$ for the predominantly semiconducting ($sem-$CNT) and metallic ($met-$CNT) films, extracted following section S2 of the supplementary material. The films show similar conductivities in terms of frequency dependence and sign, resulting from driven oscillation of the plasmon resonance at higher frequency \cite{Shuba2012,Zhang2013}. We observe a factor of three difference in conductivity between $sem$-CNTs and $met$-CNTs due to the higher free charge density in metallic tubes. 
	However, the frequency of the plasmon resonance is not expected to depend on the density of free charges \cite{DasSarma2009}. 
	When we photoexcite the films away from the optical resonances, we see that the similarity in the responses persists, as previously reported by Beard et al. \cite{Beard2008}: in figure \ref{fig:condexcvsChirality}a we plot the photoconductivity of each film, $\Delta\sigma_{ph}$, measured 1 ps after excitation ($\Delta\tau=1$ ps). This similarity suggests that the variation in ultrafast CNT photoconductivities reported in the literature \cite{Xu2009,Kampfrath2008,Beard2008,Perfetti2006,Luo2015a,Jensen2013a} is not directly linked to a variation in the chirality-distribution of the samples.
	
	\begin{figure*}[tb]
		\centering
		\includegraphics[width=1\linewidth]{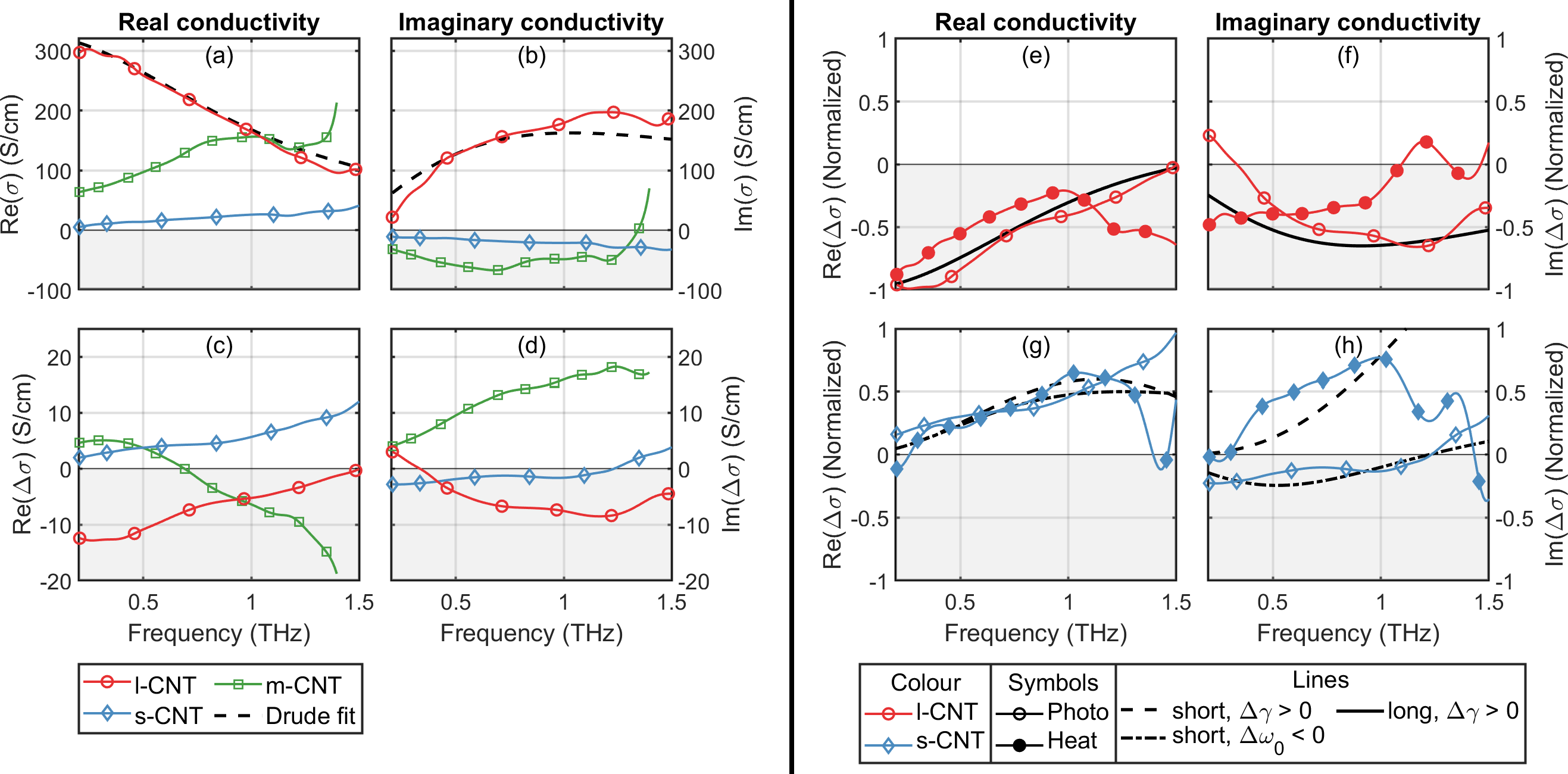}
		\caption{Effective conductivity \textbf{(a)} Re$(\sigma)$ and \textbf{(b)} Im$(\sigma)$ of $l-$, $m-$ and $s-$CNT at 300 K, as well as a Drude fit of the $l-$CNT, and change in effective conductivity \textbf{(d)} Re$(\Delta\sigma)$ and \textbf{(e)} Im$(\Delta\sigma)$ due to 800 nm photoexcitation at pump-probe delay time $\Delta\tau=1$ ps. The incident fluence is 0.7 $\mu$J/cm$^2$ for $l-$CNT, and 15 $\mu$J/cm$^2$ for $s-$ and $m-$CNTs. Note that the values of $l-$CNT have been scaled by $10^{-1}$ in \textbf{(a)}-\textbf{(d)}.
			Change in effective conductivity (Re$(\Delta\sigma)$ and Im$(\Delta\sigma)$) of \textbf{(e)-(f)} $l-$CNT and \textbf{(g)-(h)} $s-$CNT due to heating from 10 K -- 300 K (filled symbols), compared with the same $\Delta\sigma_{ph}$ data as in \textbf{(c)}-\textbf{(d)} (open symbols). Additionally, the black lines in \textbf{(e)-(h)} show the fitted change in conductivity for three simple Lorentzian resonances at resonance frequency $\omega_{0}=2\pi\times10^{-2}$ THz ($long$, full line), $\omega_{0}=2\pi\times10$ THz ($short$, dashed line), and $\omega_{0}=2\pi\times8$ THz ($short$, dash-dotted line), respectively, which have been fitted to $\Delta\sigma_{ph}$ for $l$-CNT, and $\Delta\sigma_{ph}$ and $\Delta\sigma_{heat}$ for $s$-CNT, respectively. These illustrate the difference in $\Delta\sigma$ when increasing the scattering rate $\Delta\gamma>0$ and when decreasing the resonance frequency $\Delta\omega_{0}<0$. Here we have chosen $\gamma=2\pi\times1.5$ THz and $\gamma=2\pi\times50$ THz, for the $long$ and $short$ resonances, respectively, and $\Delta\gamma=1$ THz and $\Delta\omega_0 = -1$ THz. Note that all data in \textbf{(e)}-\textbf{(h)} have been normalized by the maximum absolute value of each $\Delta\sigma$ in the displayed frequency region to make the overall frequency behaviour more comparable.}
		\label{fig:condexcvsheatvsl}
	\end{figure*}
	
	We therefore move on to consider influence of nanotube length, where we compare the $l-$, $m-$ and $s-$CNT films in figure \ref{fig:condexcvsheatvsl}, defined by average lengths 10 $\mu$m, 1 $\mu$m and $<0.3$ $\mu$m, respectively. Here, we see a drastic difference in the real and imaginary parts of $\sigma(\nu)$. For $l-$CNT, we observe a typical free electron (Drude) response (dotted line), indicating that, in this sample, the carriers are free to move along the tube length. However, the $m-$ and $s-$CNT both display a typical plasmonic resonance, located above 1.5 THz, due to the finite-length effect \cite{Slepyan2010a,Shuba2012,Karlsen2017}. Broadband infrared conductivity measurements of these CNT-films confirm that the THz resonance shifts to higher frequencies with decreasing tube length \cite{Karlsen2017}. Likewise, the photo-induced THz response is quite variable for these samples of different length CNTs, plotted in figure \ref{fig:condexcvsheatvsl}c-\ref{fig:condexcvsheatvsl}d. For short tubes ($s$-CNT), we observe a photoconductivity $\Delta\sigma_{ph}$, measured at $\Delta\tau=1$ ps, which has a positive real component and a negative imaginary component for all frequencies in our range. The medium length tubes ($m$-CNT) display a real component of photoconductivity which changes sign at approximately 0.7 THz, while the longest tubes ($l$-CNT) display a real component of the photoconductivity which is negative for all frequencies in our range.
	It is interesting to note that the photoconductivity observed for the $l-$CNT film is similar to that observed in the literature by Xu et al. \cite{Xu2009}, and the $m-$CNT is similar to the observation by Kampfrath et al. \cite{Kampfrath2008}, while the behaviour of films $s-$, $sem-$ and $met-$CNT are similar to that observed by Beard et al. and Jensen et al. \cite{Beard2008,Jensen2013a}. Thus the photoconductivities observed for our films extend across the full range of photo-responses observed previously in the literature.
	
	After illumination with a femtosecond optical stimulus, the electron  temperatures in CNTs are thought to rise by several hundred kelvin \cite{Hertel2000,Moos2002}, and the heating of electron and phonon systems occurs even at low pulse fluence, e.g. 5 $\mu$J/cm$^2$, as shown for graphite film in \cite{Kampfrath2005}. Therefore, to elucidate the origin of this rather peculiar variation in behaviour, we also investigate the change in conductivity on heating our samples. In figures \ref{fig:condexcvsheatvsl}e-\ref{fig:condexcvsheatvsl}h we compare $\Delta\sigma_{ph}$ to the change induced by heating from 10K to 300 K, ($\Delta\sigma_{heat}$). The similarity in the change in the frequency response for heating compared to photoexcitation is striking (with the exception of Im$(\Delta\sigma)$ for $s-$CNT, discussed below). Very similar behaviour is also observed for $sem-$ and $met-$CNT films, shown in supplementary material section S3. Our observations suggest that $\Delta\sigma_{ph}$ and $\Delta\sigma_{heat}$ likely originate from the same underlying mechanism, one that is related to heating induced changes in the conductivity. 
	We note that the intrinsic terahertz conductivity of metallic CNTs with diameters less than 2 nm follows the Drude law, where the plasma frequency does not depend on the temperature \cite[Equation (24)]{Slepyan1999a}.
	A dominant heating effect may arise from electron scattering by hot optical-phonons \cite{Kampfrath2005} as well as acoustical phonons, which have a linear temperature dependence of the Drude scattering rate in CNTs below 500 K \cite{Zhou2005,Karlsen2017}. 
	However, the influence that this has on the THz conductivity will depend on the frequency and oscillator strength of the THz plasmon resonance. In general, one would expect heating to induce a broadening of the THz peak. However, depending on the frequency of the resonance, this can lead to either an increase or decrease of the effective conductivity of the CNT-film in the THz range. To illustrate this effect, in figures \ref{fig:condexcvsheatvsl}e-\ref{fig:condexcvsheatvsl}h we fit the differential conductivity expected for a Lorentzian resonator, given by   
	\begin{align}
	\sigma = -i\omega\epsilon_0\left(\frac{A}{\omega^2 - \omega_0^2 + i\omega\gamma}\right),
	\end{align} 
	where $A$ is the oscillator strength, $\epsilon_0=8.85\times10^{-12}Fm^{-1}$ is the vacuum permittivity, and $\gamma$ is the scattering rate.
	We note that this simple model ignores contributions to the scattering rate from inhomogeneous broadening over CNT length \cite[Equations (1) and (2)]{Slepyan2010a}.
	The fits give us three Lorentzians with resonance frequencies located at $\omega_{0}=2\pi\times10$ THz and $\omega_{0}=2\pi\times8$ THz, representing $short$ CNTs (fitted to $\Delta\sigma_{ph}$ and $\Delta\sigma_{heat}$ for $s$-CNT, respectively), and $\omega_{0}=2\pi\times10^{-2}$ THz, representing $long$ CNTs (fitted to $\Delta\sigma_{ph}$ for $l$-CNT). 
	It is straightforward to qualitatively reproduce the general trends of the observed real part of the value $\Delta\sigma=(\partial\sigma/\partial\gamma)\Delta\gamma$ in figures \ref{fig:condexcvsheatvsl}e-\ref{fig:condexcvsheatvsl}f by assuming a heat induced increase in the scattering rate, $\gamma$. This gives rise to a change in real part of the conductivity which is negative for a low frequency resonator ($\omega_{0}=2\pi\times10^{-2}$ THz) and positive for a high frequency resonator ($\omega_{0}=2\pi\times10$ THz and $\omega_{0}=2\pi\times8$ THz).
	Based on this simple consideration we conclude that both $\Delta\sigma_{ph}$ and $\Delta\sigma_{heat}$ are determined predominantly from heat induced changes to electron scattering. 
	
	It is interesting to note that the opposite signs of Im$(\Delta\sigma_{heat})$ and Im$(\Delta\sigma_{ph})$ for $s-$CNT below 1.25 THz indicate that thermal heating and photo-excitation bring about slightly different changes to the carrier distribution. In order to reproduce this sign change, we must additionally introduce a small change to the resonance frequency, shifting to lower frequency after photoexcitation (see black dash-dotted line for $\Delta\sigma=(\partial\sigma/\partial\omega_0)\Delta\omega_0$ in figures \ref{fig:condexcvsheatvsl}g-\ref{fig:condexcvsheatvsl}h). The origin of this effect can be understood as follows: In such a percolated CNT network, the plasmon resonance frequency is determined not by the physical length of each tube, but by an effective length of conductivity pathways in the network (see ref. \cite{Karlsen2017}). On photoexcitation with optical light, some energetic carriers will be able to escape local energy minima, become more delocalized, and increase the average effective length. Such an effect will be most important for short length tubes, as observed in experiment.
	
	In conclusion, using optical pump - THz probe time-domain spectroscopy we measured the photo-induced change in THz conductivity, $\Delta\sigma_{ph}$, in free-standing carbon nanotube (CNT) films of different lengths and chirality distributions. By comparing CNT films with average individual tube lengths ranging from 0.3 $\mu$m to 10 $\mu$m, we demonstrated that drastic variations in $\Delta\sigma_{ph}$ observed for various films primarily originates from changes to the plasmonic resonance observed in finite length CNTs due to expected heat-induced changes to electron scattering. Thus we conclude that the photoexcited ultrafast THz response is predominately plasmonic in nature, and that the length of the CNTs is what determines the frequency-dependent behaviour. This explains the conflicting reports presented in \cite{Xu2009,Kampfrath2008,Beard2008,Perfetti2006,Luo2015a,Jensen2013a}, and underlines the need to carefully consider the length of the CNTs when analysing their ultrafast THz response, and more importantly, when developing nanotube-based optoelectronic devices such as photodetectors \cite{Zhang2016} and ultrafast polarization modulators \cite{Docherty2014a}, since the CNT geometry in these devices will have a huge influence on their performance. To this end, we have also shown OPTP to be a simple and efficient technique for predicting the geometry of CNT films, which currently requires careful statistical measurements with electron microscopy.
	
	This research was partially supported by the European Union’s Seventh Framework Programme (FP7) for research, technological development and demonstration under project 607521 NOTEDEV, and by EPSRC fellowship EP/K041215/1. AGN thanks the Russian Science Foundation (Project identifier: 17-19-01787) for supporting the synthesis of carbon nanotubes and film fabrication. PL, MVS and PPK thank for support H2020 project 696656 Graphene Core1, and for partial support H2020 RISE project 644076 CoExAN and H2020 RISE project 734164 Graphene 3D. PPK and MVS also are thankful for support by Tomsk State University Competitiveness
	Improvement Program.
	
	\section*{References}
	\bibliography{literature}	
\end{document}


\title[Supplementary material: Sign inversion in the terahertz photoconductivity of single-walled carbon nanotube films]{Supplementary material: Sign inversion in the terahertz photoconductivity of single-walled carbon nanotube films}
\author{Peter Karlsen}
\email{peterkarlsen88@gmail.com}
\affiliation{%
	School of Physics, University of Exeter, Stocker Road, EX4 4QL, United Kingdom
}%
\author{Mikhail V. Shuba}
\email{mikhail.shuba@gmail.com}
\author{Polina P. Kuzhir}
\affiliation{
	Institute for Nuclear Problems, Belarus State University, Bobruiskaya 11, 220050 Minsk, Belarus
}%

\affiliation{
	Tomsk State University, Lenin Avenue 36, 634050, Tomsk, Russia
}%

\author{Albert G. Nasibulin} 
\affiliation{
	Skolkovo Institute of Science and Technology, Skolkovo Innovation Center, Building 3, Moscow 143026, Russia
}%
\affiliation{
	Department of Applied Physics, Aalto University, School of Science, P.O. Box 15100, FI-00076 Espoo, Finland
}%

\author{Patrizia Lamberti} 
\affiliation{Department of Information and Electrical Engineering and Applied Mathematics, University of Salerno, Fisciano (SA), Italy
}%
\author{Euan Hendry}%
\email{E.Hendry@exeter.ac.uk}
\affiliation{%
	School of Physics, University of Exeter, Stocker Road, EX4 4QL, United Kingdom
}%
\date{\today}

\pacs{Valid PACS appear here}
\keywords{Carbon nanotubes, Time-Domain Spectroscopy, Dielectric Properties, Transfer Matrix Method}
\maketitle

\renewcommand{\thefigure}{S\arabic{figure}}
\renewcommand{\thesection}{S\arabic{section}}
\setcounter{figure}{0}

\section{Sample preparation}\label{sec:sampleprep}
To study the influence of tube length $L$ and chirality on the ultrafast photoresponse of the THz conductivity spectra we prepared five types of films comprising CNTs, where the average lengths of the CNTs  and their chirality distributions vary significantly.  The preparation methods for the $l-$, $m-$ and $s-$CNT films are identical to the ones described in ref. \cite{Karlsen2017}, and reproduced here: 

(i) Free-standing films comprising long length bundled single-walled CNTs ($l$-CNT, $L\in(2,100)$ $\mu$m, average length of 10 $\mu$m) with diameters of approximately 1.6 nm was prepared via the aerosol chemical vapour deposition method \cite{Nasibulin2011,Tian2011}. The film thickness is 55 nm and contains a mix of metallic and semiconducting CNTs, i.e. the film is not enriched. 

(ii) Free standing films comprising medium length bundled single-walled CNTs ($m$-CNT, $L\in(0.3,2)$ $\mu$m, average length of 1 $\mu$m, film thickness is 500 nm and unenriched) were prepared via the vacuum filtration technique \cite{Hennrich2002}. Briefly, a material of non-purified High Pressure Carbon Monoxide (HiPco) CNTs with diameters of 0.8--1.2 nm (NanoIntegris Inc.) were dispersed by ultrasonic treatment (Ultrasonic device UZDN-2T, 44 kHz, maximum power) for 1 hour in an aqueous suspension with 1\% Sodium-Dodecyl-Sulfate (SDS). Ultrasonic treatment cut the initially long tubes down to a length of up to 1--2 $\mu$m \cite{Hennrich2007}. Then, the suspension were centrifuged at 10000g for 15 minutes. Strong centrifugation leads to purification of the tubes and removes aggregated tubes. The suspension was then filtrated through a membrane, causing a film to collect on the filter. This film was then washed to remove all surfactant. Finally, the filter paper was dissolved by acetone and the film was transferred on to a  metallic frame with a hole of 8 mm in diameter.

(iii) Thin films of short length bundled single-walled CNTs ($s$-CNT, $L<300$ nm, film thickness is 600 nm and unenriched) were also prepared via the vacuum filtration technique. To obtain short tubes, highly purified (99\%) HiPco CNTs with diameters of 0.8--1.2 nm ({NanoIntegris} Inc.), were cut by ultrasonic treatment of the material in a mixture of nitric and sulfuric acids \cite{Shuba2012N}.    

(iv) Free standing films comprising medium length, individual, enriched semiconducting (99\%) and metallic (95\%) single-walled CNTs, respectively ($sem-$ and $met$-CNT, $L\in(0.1,1)$ $\mu$m, purchased from NanoIntegris Inc.). These samples were also prepared via the vacuum filtration technique.

\begin{figure}[tb]
	\centering
	\includegraphics[width=\linewidth]{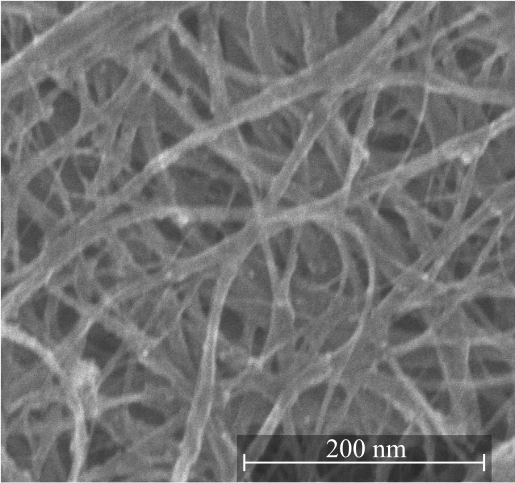}
	\caption{SEM image of a sample comprising $m$-CNTs.}
	\label{fig:SEM}
\end{figure}

To investigate the geometry of the CNT films, the $m$-CNT film was investigated using a Scanning Electron Microscope (SEM), see figure \ref{fig:SEM}, which demonstrates the successful dispersion and non-aggregation of the CNTs, since the nanotubes are dispersed evenly in the sample and no dense particles comprising ten or more nanotubes can be observed.

We note here that the thickness and diameters of the $l$-CNT films differ slightly from the rest of the films, see table 1 in the main manuscript.
A small film-thickness of 55 nm had to be chosen in order to measure in a transmission geometry, since $l$-CNTs strongly absorb THz radiation due to their high THz conductivity, see figure 3a in the main manuscript.  
The larger diameters of the $l$-CNTs is due to the difference in fabrication methods, namely the aerosol method for the $l$-CNT, and HiPco method for the rest of the films. It has been demonstrated previously that the polarizability and conductance of an individual single-walled CNT slightly depends on its diameter in the frequency range below interband transitions \cite{Slepyan2010a}. In terms of the effective conductivity of the CNT film, the tube diameter can influence this via intertube tunnelling or by varying the tube number density in the samples, however its contribution to the frequency of the terahertz plasmon resonance is much smaller than from the finite-length effect. This was shown experimentally for samples with different average tube diameters of 0.8, 1, and 1.4 nm \cite{Shuba2012}.

\section{Experimental measurement and analysis}
Throughout this paper, we choose to describe the frequency dependent electromagnetic response of the CNT-film in terms of the complex effective conductivity, $\sigma(\nu)$, as in references \cite{Beard2008,Xu2009,Zhang2013}. The complex conductivity is related to the complex permittivity, $\epsilon(\nu)$, through
\begin{equation}\label{eq:cond}
\epsilon(\nu)=1 + \frac{i\sigma(\nu)}{2\pi\nu\epsilon_0},
\end{equation}
where $\epsilon_0=8.85\times10^{-12}Fm^{-1}$ is the vacuum permittivity and $\nu$ is the frequency.
\begin{figure*}[tb]
	\centering
	\includegraphics[width=0.75\linewidth]{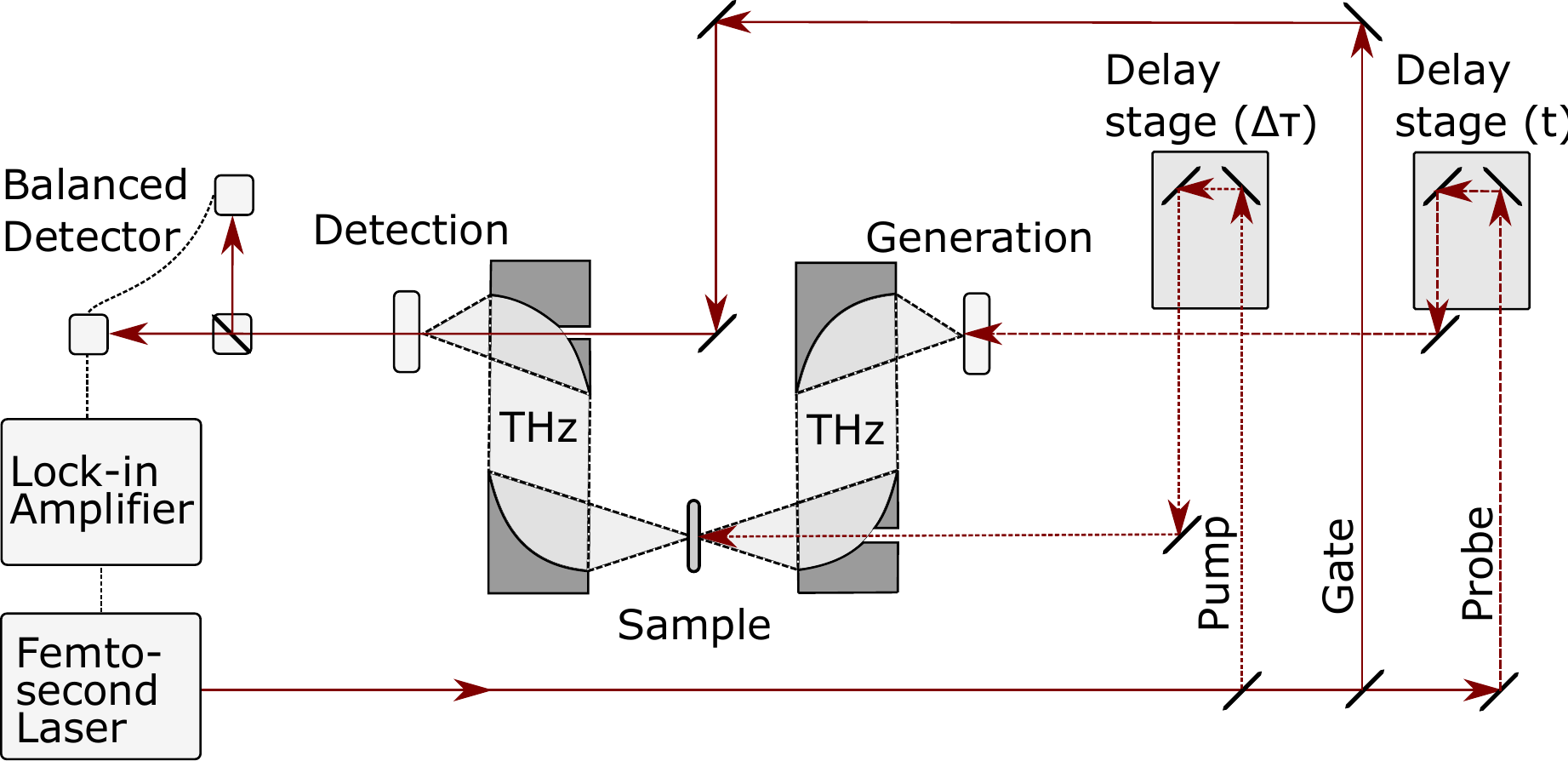}
	\caption{The optical pump - THz probe (OPTP) time domain spectrometer: Generation of THz occurs in a ZnTe crystal through optical rectification. Generated THz pulses are first collimated and then focused onto a sample using off-axis parabolic mirrors. The sample can be housed in a closed cycle helium cryostat for temperature control. Transmitted THz pulses are	re-collimated and focused onto a detection crystal. THz field waveforms are detected through electro-optical sampling in a second ZnTe crystal by scanning the temporally narrow 800 nm gate pulse over the THz field using delay stage ($t$). A pump pulse can be used to excite the sample at pump-probe delay time $\Delta\tau$ using a second delay stage ($\Delta\tau$).}
	\label{fig:OPTP}
\end{figure*}

We carried out both transmission and photoconductivity measurements over the range ~0.2--1.5 THz, where THz pulses were incident normal to our samples. Transmission spectra were obtained using a simple time-domain spectrometer, where THz pulses were generated and detected by commercially available Photoconductive Antennas (PCAs) \cite{Fattinger1989a} from \href{http://www.batop.com/index.html}{Batop} using a 40 MHz, 1064 nm, femtosecond fibrelaser from \href{http://ekspla.com/}{Ekspla}. To investigate the photoexcited THz response of our samples, we employed a 100 fs, 1050 Hz repetition rate, 800 nm Ti:Sapphire amplified laser, where THz pulses were generated and detected by optical rectification \cite{Zhang1992} and electro-optic sampling \cite{Wu1997}, respectively, in 1 mm thick ZnTe crystals, see figure \ref{fig:OPTP}. To photoexcite the sample, we again use 800 nm pulses, with fluences in the range of 0.7--15 $\mu$J/cm$^2$. To investigate the temperature dependence of the THz conductivity in the range 10--300 K, we employed a closed cycle helium cryostat (\href{http://www.arscryo.com/}{ARS})\cite{Ulbricht2011} with quartz windows.  Note that the relatively narrow bandwidth of our THz setups is determined by the transmission through this cryostat system.

\begin{figure}
	\centering
	\includegraphics[width=1\linewidth]{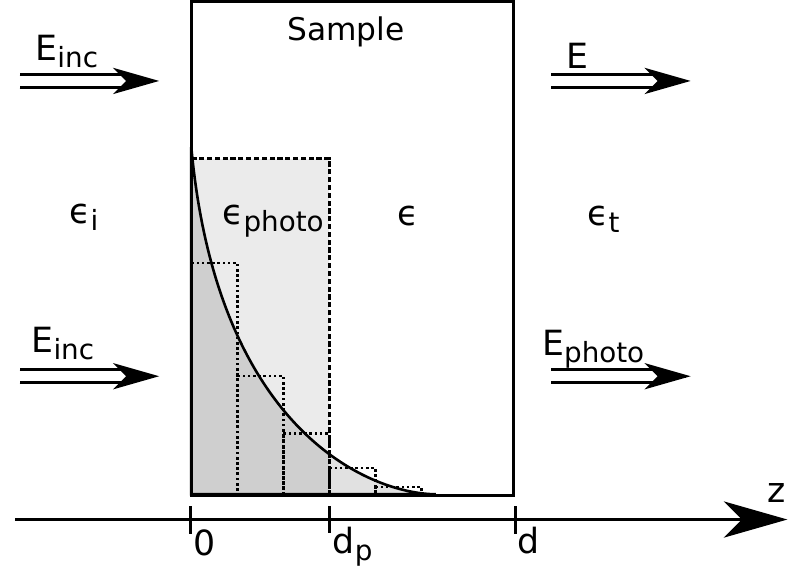}
	\caption{Diagram of the pump-probe experimental geometry, showing an incident probe-pulse $E_{inc}$ on a sample with permittivity $\epsilon$ and thickness $L$, surrounded by an incident region $\epsilon_i$ and transmitted region $\epsilon_t$. Two measurements are obtained in this case: transmission through the photoexcited sample $E_{photo}$, and a corresponding reference, $E$, through the unexcited sample. The photoinduced response $\Delta\epsilon(z)=\Delta\epsilon_0 e^{-z/d_p}$ decays exponentially within the sample due to attenuation of the incident pump, where $\Delta\epsilon_0$ is the photo-induced change in the permittivity at the surface of the sample, and $d_p$ is the distance at which $\Delta\epsilon$ has decayed to $e^{-1}$. We model this by dividing the sample into a number of homogeneous layers, $N_L$, each with a permittivity $\epsilon_{photo}(z)=\epsilon+\Delta\epsilon(z)$.}
	\label{fig:photoexc2}
\end{figure}

By analysing the frequency-dependent transmission amplitude and phase, $t(\nu)$, of a sample, we can determine its complex effective conductivity, $\sigma(\nu)$, as in references \cite{Beard2008,Xu2009,Zhang2013}, where $\nu$ is the frequency. Similarly, by recording the difference in transmission, $\Delta{}E=E_{ph}-E$, between a photoexcited ($E_{ph}$) and unexcited sample ($E$), a complex photoconductivity $\Delta\sigma_{ph}(\nu,\Delta\tau)$ can be obtained as a function of pump-probe delay-time $\Delta\tau$. However, in order to simplify the notation of our equations in this section, we write them in terms of the complex permittivity $\epsilon$ and photo-induced change in the permittivity $\Delta\epsilon$ using equation \eqref{eq:cond}. The geometry of our photoexcited sample is shown in figure \ref{fig:photoexc2}, where the sample is surrounded by an incident region ($\epsilon_i$) and a transmitted region ($\epsilon_t$). Photoexcitation of the sample results in a change in the permittivity $\Delta\epsilon=\epsilon_{ph}-\epsilon$, where $\epsilon_{ph}$ is the permittivity of the photoexcited sample. Typically $\Delta\epsilon$ will decay exponentially within the sample due to linear attenuation of the incident pump-pulse, i.e. $\Delta\epsilon(z)=\Delta\epsilon_0e^{-z/d_p}$, where $\Delta\epsilon_0$ is the photo-induced change in permittivity at the surface of the sample, and $d_p$ is the penetration depth of the pump pulse at which $\Delta\epsilon(z)$ has decayed by a factor $e^{-1}$. Here $d_p$ is determined from transmittance measurements using the Beer-Lambert law $I_t=I_0e^{-d/d_p}$, where $I_0$ and $I_t$ is the  incident and transmitted pump intensity, respectively, and $d$ is the thickness of the sample. The photoexcited sample can be represented by dividing it into $N_{L}$ homogeneous layers with photoexcited permittivities $\epsilon_{ph}(z)=\epsilon+\Delta\epsilon(z)$. By way of the transfer matrix method this multilayer system can then be represented in the following way \cite{Ulbricht2011}:
\begin{align}
\begin{pmatrix}
E_{01} \\
Z_0 H_{01}
\end{pmatrix} &= M_{tot}
\begin{pmatrix}
E_{N,N+1} \\
Z_0 H_{N,N+1}
\end{pmatrix}, \label{eq:Mtotal}
\end{align}
where $E_{j,j+1}$ and $H_{j,j+1}$ is the electric and magnetic field, respectively, at the interface between layer $j$ and $j+1$, $Z_0$ is the free-space impedance and $M_{tot}$ is total matrix of the $N$-layer system:
\begin{align}
M_{tot}&=M_1 M_2 \cdots M_N=\begin{pmatrix}
m_{11} & m_{12} \\
m_{21} & m_{22}
\end{pmatrix}, \\
\text{where} \quad M_j &= 
\begin{pmatrix}
\cos{\delta_j} & \frac{-i \sin{\delta_j}}{\gamma_j} \\
-i\gamma_j\sin{\delta_j} & \cos{\delta_j}
\end{pmatrix}.
\end{align}
Here $M_j$ is the matrix associated with layer $j$, $\delta_j = 2\pi\nu\sqrt{\epsilon_j} d_j \cos{\theta_j}/c$ is the phase delay through layer $j$, and $\gamma^{TE}_j = \sqrt{\epsilon_j}\cos{\theta_j}$ and $\gamma^{TM}_j = \cos{\theta_j}/\sqrt{\epsilon_j}$ for TE- and TM-polarization, respectively. The transmission coefficient of the multilayer system is then given by
\begin{align}
t&=\frac{2\gamma_i}{\gamma_i m_{11} + \gamma_i\gamma_t m_{12} + m_{21} + \gamma_t m_{22}}. \label{eq:transmission}
\end{align}

By measuring the change in transmitted electric field through the sample due to photoexcitation $\Delta{}E=E_{ph}-E$ at pump-probe delay time $\Delta\tau$, one can obtain the transmission coefficient of the photoexcited region in question:
\begin{align}\label{eq:tn}
\frac{\Delta{}E(\Delta\epsilon_0,\Delta\tau,\nu)}{E(\nu)} = \frac{t_{ph}(\Delta\epsilon_0,\Delta\tau,\nu)-t(\nu)}{t(\nu)},
\end{align} 
which can either be solved numerically for the unknown $\Delta\epsilon_0$ in its current form, or approximated further, given that certain conditions about the geometry and wavelength are true \cite{Kuzel2007,Nienhuys2005e}. We choose to solve equation \eqref{eq:tn} numerically using $N_L=5$, since some of our samples are in an intermediate regime in terms of the sample thickness, pump penetration depth and wavelength, where none of the approximations are entirely appropriate. From the extracted $\Delta\epsilon_0$, the photoconductivity $\Delta\sigma_{ph}$ is then obtained using equation \eqref{eq:cond}. 

\section{Photoconductivity of the CNT-films}
\begin{figure*}
	\centering
	\includegraphics[width=\linewidth]{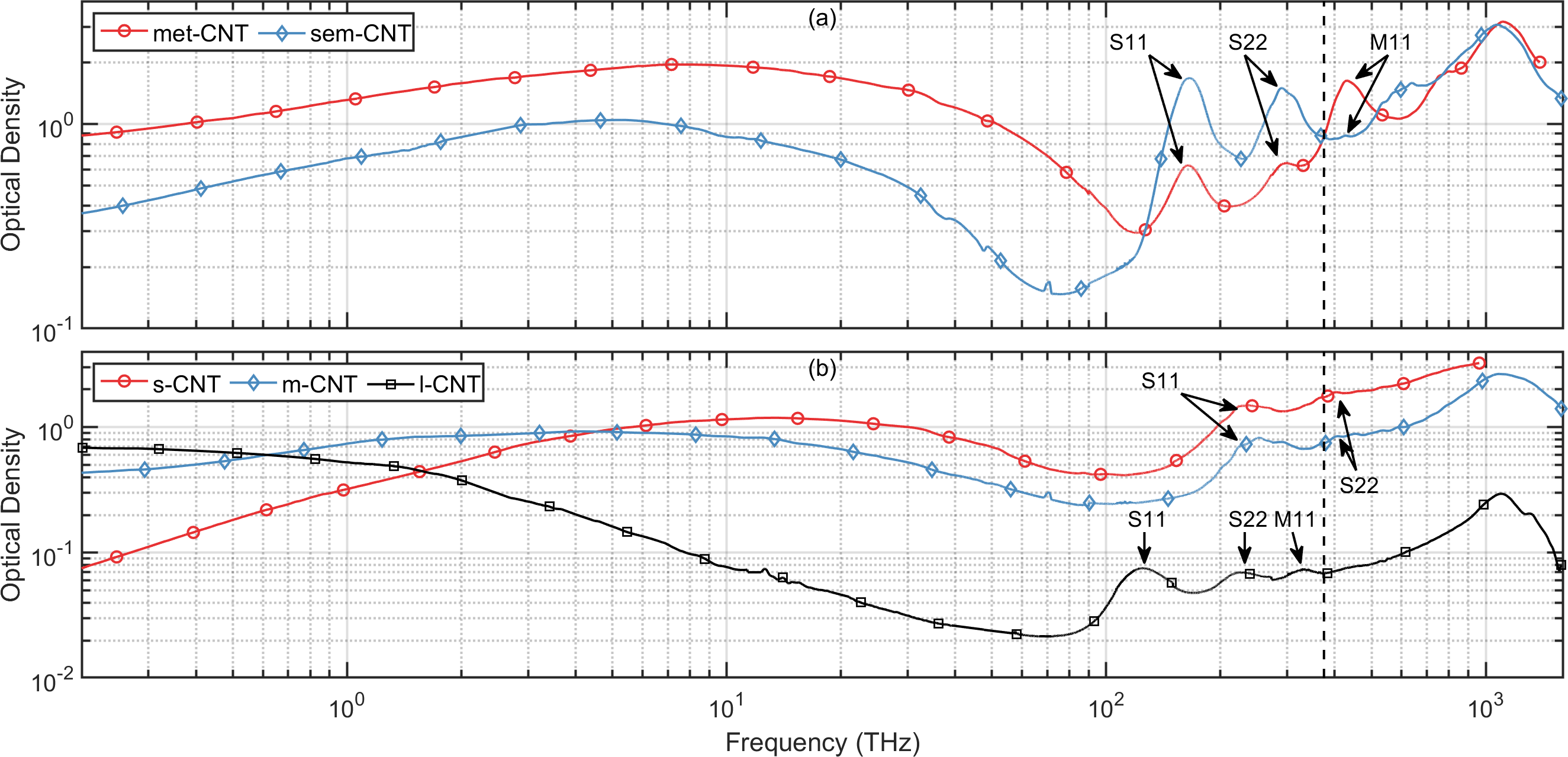}
	\caption{Optical density of \textbf{(a)} $met-$ and $sem-$CNT and \textbf{(b)} $s-$, $m-$ and $l-$CNT, measured at 323 K. The vertical dashed line corresponds to the mean frequency of the 800 nm pump pulse, and the arrows indicate the first (S11) and second (S22) optical transitions in the semiconducting CNTs and the first optical transition in the metallic CNTs (M11).}
	\label{fig:OD}
\end{figure*}
In figure \ref{fig:OD} we plot the broadband optical density spectra of the CNT films, obtained following the procedure described in reference \cite{Karlsen2017}. We note that the 800 nm photoexcitation occurs off-resonance from the optical transitions in the CNTs, specifically the first (S11) and second (S22) optical transition in semiconducting CNTs, and the first optical transition in metallic CNTs (M11). In case of the $s-$ and $m-$CNTs, the 800 nm pump overlaps with the edge of the S22 transition, meaning only a small subset of the semiconducting CNTs are photoexcited on-resonance. Therefore we do not expect these ``on-resonance" semiconducting CNTs to contribute significantly to the overall photoconductivity, compared to the rest of the CNTs.

\begin{figure}
	\centering
	\includegraphics[width=1\linewidth]{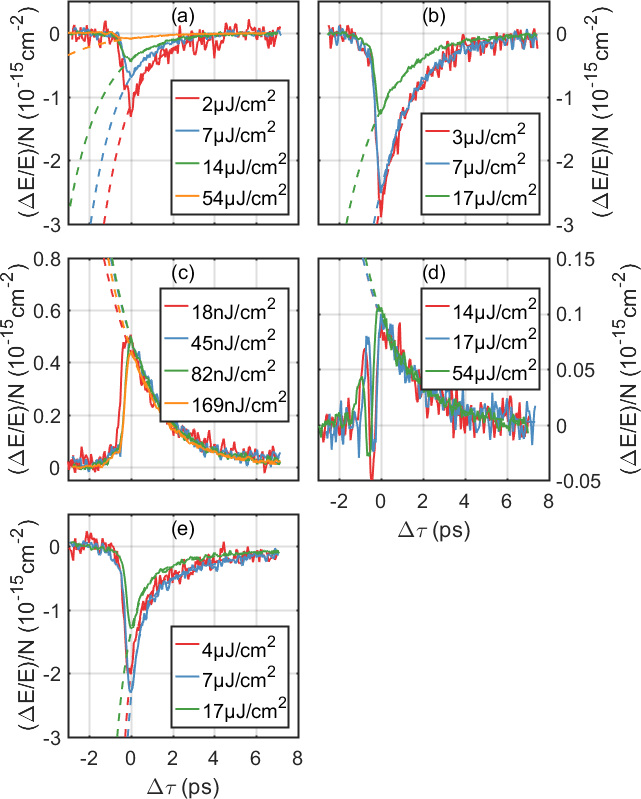}
	\caption{The photo-induced relative change in the THz transmission $\Delta{}E/E$ due to 800 nm photoexcitation at 300 K of \textbf{(a)} $met-$CNT, \textbf{(b)} $sem-$CNT, \textbf{(c)} $l-$CNT, \textbf{(d)} $m-$CNT, and \textbf{(e)} $s-$CNT, vs pump-probe delay time $\Delta\tau$ and fluence, and normalized by the absorbed photon density $N$. The solid lines are the experimentally obtained data, and the dashed lines are exponential fits. The decay time for each CNT film is \textbf{(a)} 1.3-2.2 ps, \textbf{(b)} 1.6-1.8 ps, \textbf{(c)} 1.9-2.2 ps, \textbf{(d)} 1.8-1.9 ps, and \textbf{(e)} 0.5-0.6 ps and 3.5-4.4 ps for the fast and slow decay, respectively, where the decay time is observed to slightly increase with increasing fluence.}
	\label{fig:pump_vs_fluence}
\end{figure}
In figure \ref{fig:pump_vs_fluence} we plot the photo-induced relative change in THz transmission $\Delta{}E/E$, normalized to the absorbed photon density $N$, of the CNT films described in section \ref{sec:sampleprep} for incident fluences in the range of approximately 0.2--54 $\mu$J/cm$^2$. 
For the majority of the films, the relaxation dynamics and the normalized photoresponse remains unchanged, except for relatively high fluences, where the signal becomes saturated and the decay time is observed to increase slightly. The exception is $met$-CNT in figure \ref{fig:pump_vs_fluence}a, where we observe an increased saturation of the signal for all incident fluences, which is likely due to the high charge-carrier concentration in these metallic CNTs.

\begin{figure*}
	\centering
	\includegraphics[width=1\linewidth]{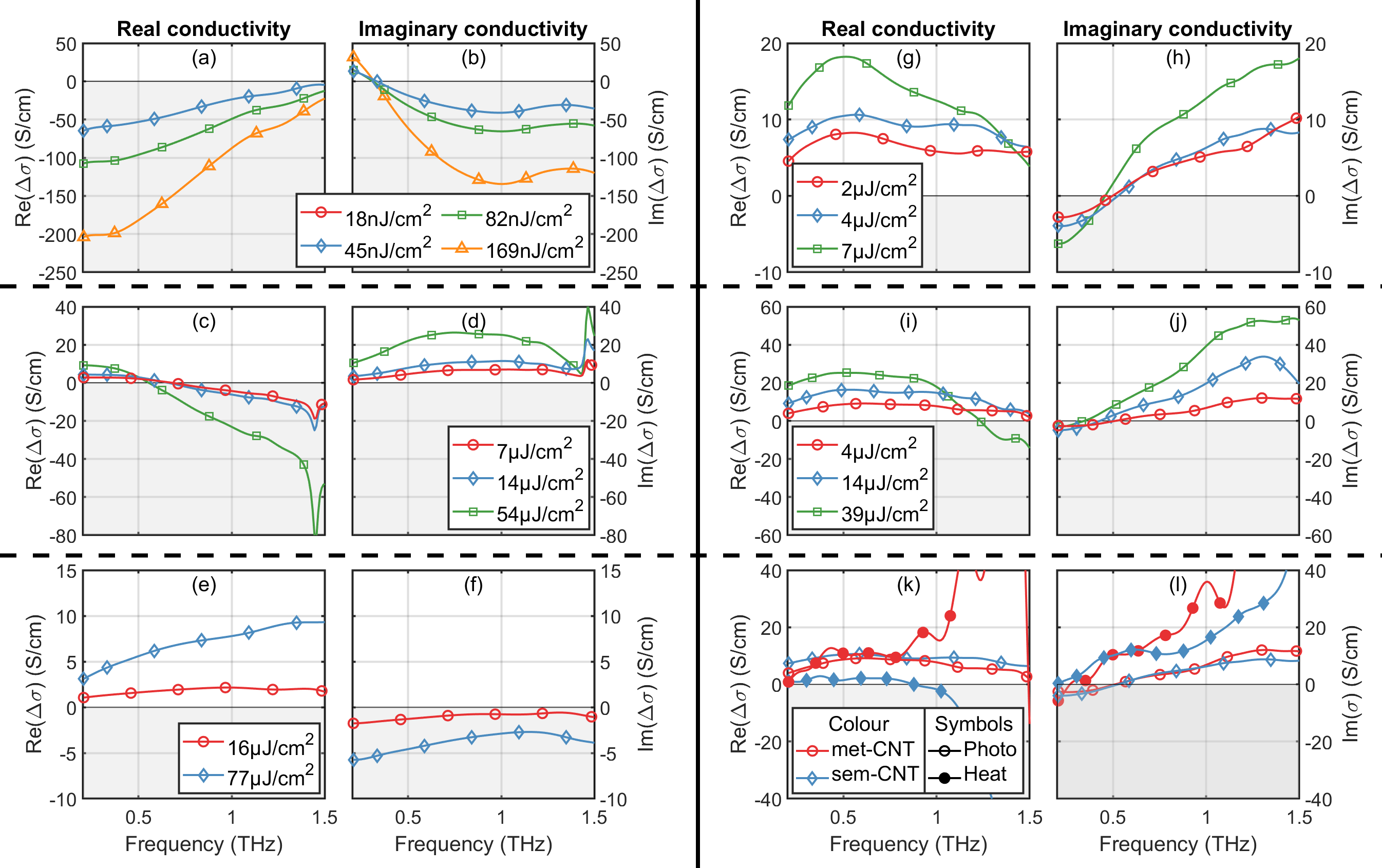}
	\caption{Change in effective conductivity Re$(\Delta\sigma)$ and Im$(\Delta\sigma)$ vs fluence for \textbf{(a)}-\textbf{(b)} $l-$CNT, \textbf{(c)}-\textbf{(d)} $m-$CNT, \textbf{(e)}-\textbf{(f)} $s-$CNT, \textbf{(g)}-\textbf{(h)} $met-$CNT, and \textbf{(i)}-\textbf{(j)} $sem-$CNT, due to 800 nm photoexcitation at pump-probe delay time $\Delta\tau=1$ ps.
		Change in effective conductivity \textbf{(k)} Re$(\Delta\sigma)$ and \textbf{(l)} Im$(\Delta\sigma)$ of $met-$ and $sem-$CNT due to heating from 10 K -- 300 K (filled symbols), compared with the same $\Delta\sigma_{ph}$ data as in \textbf{(g)}-\textbf{(j)} for incident fluence 4 $\mu{}J/cm^2$ (open symbols).}
	\label{fig:cond_vs_fluence}
\end{figure*}
In figures \ref{fig:cond_vs_fluence}a-\ref{fig:cond_vs_fluence}j we plot Re($\Delta\sigma_{ph}$) and Im($\Delta\sigma_{ph}$) of our CNT films for different incident fluences, and we observe that the responses of the CNT films are relatively fluence-independent in terms of the overall frequency dependence of $\Delta\sigma_{ph}$. In figures \ref{fig:cond_vs_fluence}k-\ref{fig:cond_vs_fluence}l we compare $\Delta\sigma_{ph}$ with the heating-induced change from 10 K to 300 K ($\Delta\sigma_{heat}$) for $met-$ and $sem-$CNT and show that the overall frequency response is almost identical for $\Delta\sigma_{ph}$ and $\Delta\sigma_{heat}$, similar to what was observed for $l-$ and $s-$CNT in the main text. The main difference between $sem-$ and $met-$CNT is a four times increase in the magnitude of Re$(\Delta\sigma_{heat})$ between $met-$ and $sem-$CNT, which is expected due to bandgap of the latter.
We note that the $s-$CNT data plotted in this section was obtained from a film placed on a quartz substrate, as opposed to the identical free-standing film described in section \ref{sec:sampleprep} and the main text, however the observed dynamics of two films are identical.

Finally, we note that CNT conductivity can be affected by doping from adsorbed molecules. As was recently shown \cite[Table 1]{Gorshunov2018}, strong doping leads to a decrease of the Drude relaxation rate in CNTs by 15\% and 24\% for iodine and CuCl doping respectively. In our samples, some molecules may have attached to the CNTs during the cutting (for $s$-CNTs) or separation processes (for $sem$- and $met$-CNTs), which in turn may lead to a doping effect \cite{Itkis2002}. To remove these impurities, all the CNT films were annealed at 270 $^\circ$C for 15 minutes. However, unintentional doping of all our samples by air molecules may still be present. Prominent absorption bands corresponding to the first interband transition in semiconducting tubes (S11 in figure \ref{fig:OD}) indicate a quite low extent of CNT doping in all our samples. The difference in the terahertz plasmon frequencies between $s$-, $m$-, and $l$-CNT samples is hundreds of percent (see terahertz absorbance bands in figure \ref{fig:OD}), while the possible changes in the Drude relaxation rate due to the doping is expected to be less than 24\% \cite{Gorshunov2018}. Therefore, we do not believe that unintentional doping in our samples could affect the sign of the terahertz photoresponse, which is the main topic of this paper.

\clearpage
\section*{References}
\bibliography{literature}